\documentclass[prl,twocolumn]{revtex4-1}
\pdfoutput=1

\usepackage{amsmath,amssymb,graphicx}

\begin{document}

\newcommand{\Br}{{\bf r}}
\newcommand{\Bu}{{\bf u}}
\newcommand{\Ba}{{\bf a}}
\newcommand{\Bs}{{\bf s}}
\newcommand{\Bsz}{{\bf s_0}}
\newcommand{\Brho}{\mbox{\boldmath $\rho$}}
\newcommand{\Bepsilon}{\mbox{\boldmath $\epsilon$}}
\newcommand{\Bkappa}{\mbox{\boldmath $\kappa$}}
\newcommand{\Bsrho}{\mbox{\boldmath $\scriptstyle \rho$}}
\newcommand{\BK}{{\bf K}}
\newcommand{\Fs}{\mathcal{F}}
\newtheorem{fig}{Fig.}
\newcommand{\dint}{\int \!\!\! \int}
\newcommand{\dist}{\sigma}
\newcommand{\dv}{{\rm d}}
\newcommand{\ii}{{\rm i}}
\newcommand{\dpone}{{\dv^2\rho_1'}}
\newcommand{\dptwo}{{\dv^2\rho_2'}}
\newcommand{\dts}{{\dv^2s}}
\newcommand{\gta}{\raisebox{-.6ex}{$\stackrel{\textstyle >}{\sim}$}}
\newcommand{\etal}{\emph{et al.}}
\renewcommand{\Im}{\mbox{Im}}
\renewcommand{\Re}{\mbox{Re}}
\newcommand{\Wcoh}{W^{\mbox{\footnotesize coh}}}
\newcommand{\xhat}{{\bf \hat{x}}}
\newcommand{\yhat}{{\bf \hat{y}}}
\newcommand{\zhat}{{\bf \hat{z}}}
\newcommand{\Arg}{\mbox{Arg}}

\newcommand{\Brhohat}{\mbox{\boldmath $\hat{\rho}$}}

\newcommand{\Bshat}{{\bf \hat{s}}}

\newcommand{\BT}{{\bf T}}
\newcommand{\BE}{{\bf E}}
\newcommand{\BF}{{\bf F}}
\newcommand{\BB}{{\bf B}}
\newcommand{\BI}{{\bf I}}
\newcommand{\BL}{{\bf L}}
\newcommand{\BM}{{\bf M}}
\newcommand{\BH}{{\bf H}}
\newcommand{\BR}{{\bf R}}
\newcommand{\BS}{{\bf S}}
\newcommand{\BV}{{\bf V}}

\newcommand{\Bv}{{\bf v}}
\newcommand{\Bl}{{\bf l}}

\newcommand{\m}{\mbox{ m}}
\newcommand{\um}{\,\mu\mbox{m}}
\newcommand{\mm}{\mbox{ mm}}
\newcommand{\nm}{\mbox{ nm}}
\newcommand{\cm}{\mbox{ cm}}

\newcommand{\pp}[2]{\frac{\partial #1}{\partial #2}}

\newcommand{\Btau}{\mbox{\boldmath $\tau$}}

\title{Fractional vortex Hilbert's Hotel}

\author{Greg Gbur}
\affiliation{Department of Physics and Optical Science, UNC Charlotte, Charlotte, NC, 28223}

\begin{abstract}
We demonstrate how the unusual mathematics of transfinite numbers, in particular a nearly perfect realization of Hilbert's famous hotel paradox, manifests in the propagation of light through fractional vortex plates.  It is shown how a fractional vortex plate can be used, in principle, to create any number of ``open rooms,'' i.e. topological charges, simultaneously.  Fractional vortex plates are therefore demonstrated to create a singularity of topological charge, in which the vortex state is completely undefined and in fact arbitrary.
\end{abstract}


\maketitle

It seems to be an unspoken adage of theoretical physics that all fields of mathematics, no matter how abstract, paradoxical or seemingly divorced from reality, inevitably find their realization or application in physical systems.  One of the strangest such fields, which until recently seemed somewhat immune to this adage, is the study of transfinite numbers, originally investigated by Cantor \cite{gc:cttfottotn}. The smallest transfinite number is the size of the set of natural numbers, typically labeled by $\aleph_0$.  In Cantor's analysis, every infinite set that can be put into one-to-one correspondence with the natural numbers is equivalent, making statements such as $\aleph_0+1=\aleph_0$ and $\aleph_0+N=\aleph_0$ quantitative (for more details see, for instance, \cite{jb:its}).  

A demonstration of this strangeness is known as ``Hilbert's Hotel,'' originally attributed to David Hilbert in a 1924 lecture but popularized by George Gamow some years later \cite{gg:otti}.  We imagine a hotel with a countably infinite number of rooms and no vacancies, with rooms labeled $1,2,3,\ldots$. Though the hotel is completely filled, it is always possible to add a new guest by moving every current guest to the next highest-numbered room.  This can be done to free up any finite number of rooms, and indeed can even be done to accommodate a countably infinite number of new guests.

In recent years, it has been demonstrated that this mapping can be achieved in quantum mechanical systems with a countably infinite number of modes.  A system which can accommodate a single new ``guest'' was introduced by Oi et al. \cite{dklovpjj:prl:2013} in the context of cavity QED, in which all quantum amplitudes are shifted up a level, leaving an unoccupied vacuum state. More recently, Poto{\u{c}}ek et al. \cite{vpfmmmmosmlacldklorwbjj:prl:2015} demonstrated a quantum-optical system that maps each state to a state with twice the original quantum number, thus realizing a Hilbert Hotel with an infinite number of new guests.

However, an even more overt realization of Hilbert's Hotel can be realized with an entirely classical field.  A decade ago, it was theoretically postulated \cite{mvb:joa} and experimentally observed \cite{jleymjp:njp:2004} that an optical beam passing through a half-integer spiral phase plate produces a chain of optical vortex pairs that is, in principle, infinite.  In this paper we demonstrate that this chain mimics exactly Hilbert's Hotel, and that in fact the mathematics of transfinite numbers are in fact a key ingredient in the behavior of the system; this relationship does not appear to have been previously recognized.  Furthermore, we extend the original example to demonstrate that it is possible to simultaneously incorporate any finite number of additional vortex ``guests'' in this system with a straightforward modification.

\begin{figure*}
\centering
\includegraphics[scale=.6]{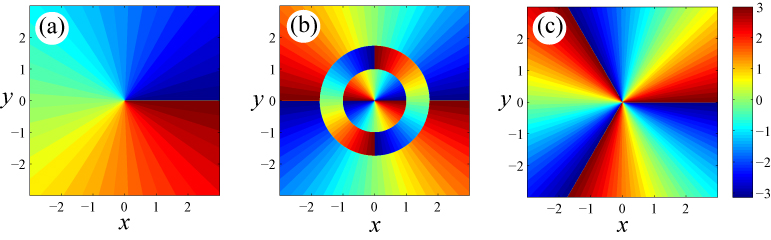}
\caption{The phase of Laguerre-Gauss beams in the waist plane, for orders (a) $n=0$, $m=1$, (b) $n=2$, $m=2$ (c) $n=0$, $m=-3$.}
\label{fig:laguerregauss}
\end{figure*}

The study of phase singularities in optical wavefields has grown over the past few decades into its own vibrant subfield of optics known as singular optics \cite{mssmvv:pio,mrdkohmjp:pio}.  The singularity typically manifests as a line of zero intensity in three-dimensional space, around which the phase has a circulating or helical structure.  The simplest examples of such singularities appear in monochromatic paraxial Laguerre-Gauss beams; those beams of non-zero azimuthal order $m$ have line singularities on their propagation axis. The phase in the waist plane of several typical beams is shown in Fig.~\ref{fig:laguerregauss}.  The phase singularity can be identified as the point at which all colors (phases) meet.

It is to be noted that the phase increases or decreases by an integer multiple of $2\pi$ in a closed circuit around the singularity.  The number of multiples is known as the \emph{topological charge} $t$ of the vortex, and the total charge within a closed path $C$ may be determined by an integral of the gradient of the wavefield phase $\psi(\Br)$, i.e.
\begin{equation}
t \equiv \frac{1}{2\pi}\oint_C \nabla \psi(\Br)\cdot d\Br.\label{topchg}
\end{equation}
The topological charge is a conserved quantity, and therefore under most circumstances vortices can only be created or destroyed in pairs of opposite sign; we will see here how Hilbert's Hotel provides an alternative.

\begin{figure}
\centering
\includegraphics[scale=.7]{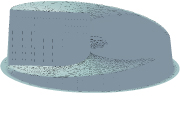}
\caption{Illustration of a spiral phase plate.}
\label{fig:spiralplate}
\end{figure}

The earliest experiments on vortex beams typically generated them using a spiral phase plate consisting of a ramp of dielectric material \cite{mwbrpccmkjpw:oc:1994}, as illustrated in Fig.~\ref{fig:spiralplate}.  Assuming geometric propagation through the material, plates can be designed to have a transmission function $t(\phi)= \exp[im\phi]$, with $\phi$ the azimuthal angle and $m$ an integer, therefore imparting the needed phase twist on the beam.  There is no prohibition, however, in fabricating a phase plate that produces a fractional twist $\alpha$; in such a case, what is the behavior of the transmitted field?

Following Berry \cite{mvb:joa}, from which many of the following propagation formulas are derived, we assume a monochromatic scalar plane wave of unit amplitude and wavenumber $k$ normally incident on a phase plate with transmission function
\begin{equation}
t(\phi) = e^{i\alpha \phi},
\end{equation}
where $\alpha$ may be positive, negative, or fractional.  Further, we neglect the contribution of evanescent waves and restrict ourselves to Fresnel diffraction.  For an integer step $\pm n$, $n\geq 0$, it can be shown that the field is of the form
\begin{eqnarray}
&&U_n(\rho,\phi,z) = e^{ikz}e^{\pm in\phi}e^{ik\rho^2/4z}\sqrt{\frac{\pi}{8}}(-i)^{n/2}\sqrt{\frac{k\rho^2}{z}} \nonumber \\ &\times&\left[ J_{(n-1)/2}(k\rho^2/4z)-i J_{(n+1)/2}(k\rho^2/4z)\right],\label{Un:final}
\end{eqnarray}
where $z$ is the distance from the phase plate.  To determine the field of a fractional phase plate, we Fourier expand the fractional transmission function in the form,
\begin{equation}
e^{i\alpha \phi} = \frac{e^{i\pi \alpha}\sin(\pi \alpha)}{\pi}\sum_{n=-\infty}^\infty \frac{e^{in\phi}}{\alpha - n}.
\end{equation}
This leads to a field of the form
\begin{equation}
U_\alpha(\Br) = \frac{e^{i\pi \alpha}\sin(\pi \alpha)}{\pi}\sum_{n=-\infty}^\infty \frac{U_n(\Br)}{\alpha - n}.
\end{equation}

\begin{figure*}
\centering
\includegraphics[scale=.4]{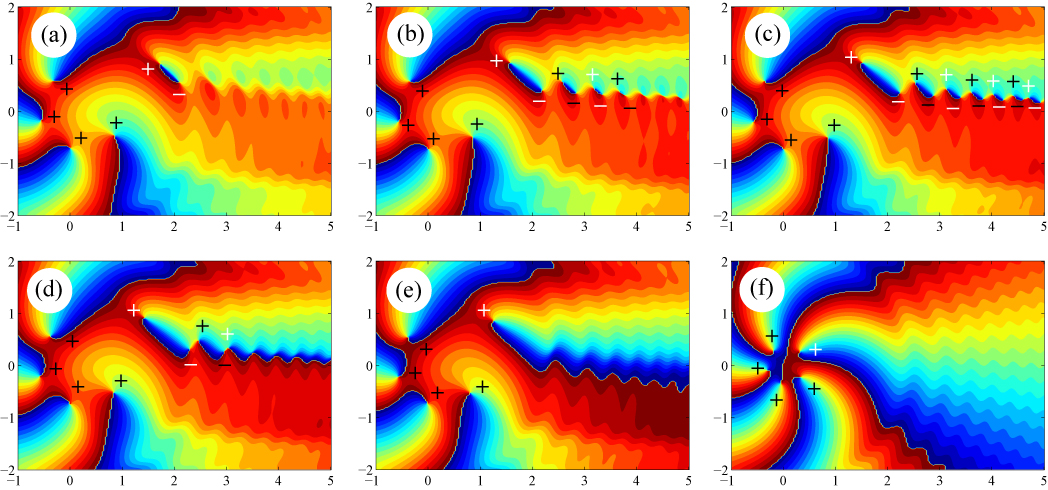}
\caption{Evolution of the field transmitted through a fractional plate, for (a) $\alpha = 4.4$, (b) $\alpha = 4.47$, (c) $\alpha = 4.5$, (d) $\alpha = 4.55$, (e) $\alpha = 4.65$, (f) $\alpha = 4.995$.  For convenience, adjacent charges that were created together are shown in the same color.}
\label{fig:vortexhotel}
\end{figure*}

We look at the evolution of the phase of the field as $\alpha$ changes from $\alpha = 4$ to $\alpha = 5$ in Fig.~\ref{fig:vortexhotel}.  The plot is done in the scaled variables $\xi = \sqrt{k/2z}x$ and $\eta = \sqrt{k/2z}y$.  As $\alpha$ approaches $\alpha=4.5$, a line of vortices are pair produced along the phase discontinuity.  The first vortices appear close to the central axis, but new pairs are rapidly produced at increasingly larger distances.  At $\alpha = 4.5$, there are an infinite number of pairs along this line, as demonstrated by Berry \cite{mvb:joa}.  As $\alpha$ increases past $\alpha = 4.5$, the singularities annihilate from the most distant points towards the origin, but with their opposite neighbor, instead of their original pair member.

It is this process that represents, in a strikingly exact way, the phenomenon of Hilbert's Hotel.  For $\alpha<4.5$, the topological charge $t=4$; in order to change to $t=5$ for $\alpha>4.5$, an unbalanced charge must come from somewhere, though it appears to be forbidden due to topological charge conservation.  The system resolves this by creating a countably infinite set of pairs of vortices.  Let us imagine that each positive charge represents a ``room'' and each negative charge a ``guest.'' Each guest has stepped out of each room (through pair creation), and then moves to the room on the right (pair annihilation).  The net result is a single additional unbalanced positive charge or, in terms of Hilbert's Hotel, a single additional unoccupied room.

\begin{figure}
\centering
\includegraphics[scale=.6]{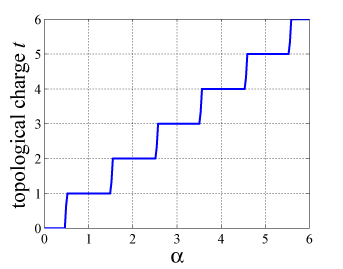}
\caption{Topological charge as a function of $\alpha$, calculated by numerically evaluating the integral of Eq.~(\ref{topchg}) at $\sqrt{\xi^2+\eta^2}=20$.}
\label{fig:topchg}
\end{figure}

It can be said that, at least for this particular system configuration, a new charge is created by creating a true singularity of topological charge.  When there are an infinite number of pairs, the topological charge of the field is completely undefined, as in principle any number of unbalanced charges could be taken from the line and still have all remaining pairs annihilate. For this system, then, new charge is created by applying transfinite arithmetic.  A plot of topological charge as a function of $\alpha$ is shown in Fig.~\ref{fig:topchg}.

\begin{figure}
\centering
\includegraphics[scale=.6]{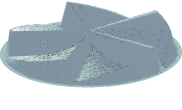}
\caption{Illustration of a multi-ramp phase plate, with $m=5$.}
\label{fig:multiramp}
\end{figure}

We may extend this result to create an arbitrary number of vortices in a single step, just as any finite number of guests may be accommodated at once in the Hotel.  We consider the transmission function given by
\begin{equation}
t(\phi) = e^{i\alpha(\phi-2\pi k/m)}, \quad 2\pi k/m \leq \phi < 2\pi(k+1)/m.
\end{equation}
with $k = 0,1,\ldots, m-1$. This represents a multi-ramp phase plate, as illustrated in Fig.~\ref{fig:multiramp}.  

\begin{figure*}
\centering
\includegraphics[scale=.6]{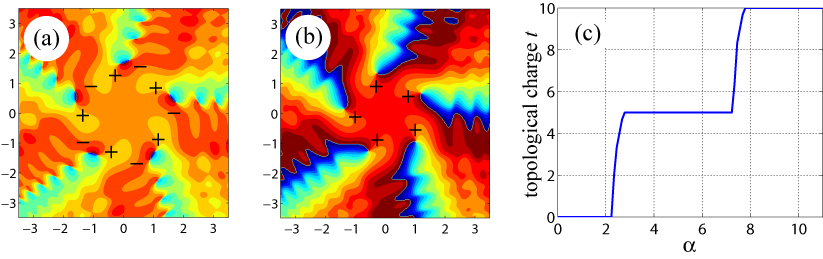}
\caption{Illustrating a jump of topological charge greater than one for $m=5$. (a) The case $\alpha = 1.95$. (b) The case $\alpha = 3.2$. (c) The topological charge as a function of $\alpha$.}
\label{fig:multichargejump}
\end{figure*}

It can be shown that the Fourier series coefficients of this transmission function vanish for any index not a multiple of $m$; for $n=qm$, with $q=\ldots, -2,-1,0,1,2,\ldots$, we have
\begin{equation}
c_{qm} = \frac{\sin(\alpha \pi/m)}{\pi(\alpha/m-q)}e^{i\alpha \pi/m}.
\end{equation}
Now, as $\alpha$ approaches the value $m/2$, $m$ lines of vortices are created along each discontinuity of the transmission function, as can be seen in Fig.~\ref{fig:multichargejump}(a).  The same annihilation process happens after $\alpha=m/2$, leaving an unbalanced $m$ charges, as seen in Fig.~\ref{fig:multichargejump}(b). The topological charge then jumps from 0 to $m$ at once, as illustrated in Fig.~\ref{fig:multichargejump}(c).

The discussion of an infinite set of objects in any physical system must, of course, come with significant caveats.  We have used a plane wave of infinite transverse extent in this discussion, while a realistic optical field must have finite width.  It is possible, however, to calculate the propagation of a Gaussian beam with field $U_0(\Br') = \exp[-r'^2/2w_0^2]$ directly through Fresnel propagation; the formula for $U_n(\Br)$ that results is
\begin{eqnarray}
 U_n&&\!\!\!\!\!\!\!(\rho,\phi,z) = e^{ikz}e^{\pm in\phi}e^{ik\rho^2/4z(1-R/2)}\sqrt{\frac{\pi}{8}}(-i)^{n/2}\sqrt{\frac{k\rho^2R}{z}} \nonumber \\ &\times&R\left[ J_{(n-1)/2}(k\rho^2R/4z)-i J_{(n+1)/2}(k\rho^2R/4z)\right],
\end{eqnarray}
with $R(z)=1/(1+iz/kw_0^2)$.  The formula is nearly identical to that of Eq.~(\ref{Un:final}) earlier, with only the addition of the propagation factor $R(z)$.  Provided $z\ll kw_0^2$, or we restrict ourselves to propagation distances smaller than the Rayleigh range, the finite beam should well approximate the infinite plane wave.

\begin{figure*}
\centering
\includegraphics[scale=.5]{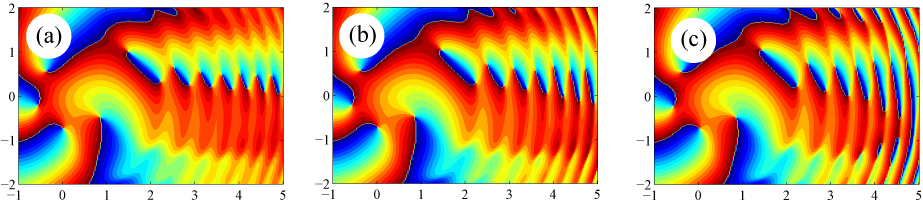}
\caption{Illustrating the ``vortex Hotel'' chain for a beam at different propagation distances, with (a) $z = 0.5\m$, (b) $z=0.7\m$ (c) $z=0.9\m$, with $w_0 = 1\mm$, $\lambda = 500 \nm$.  Here $\alpha = 4.47$.}
\label{fig:beamfractional}
\end{figure*}

An illustration of the vortex chain in beams is shown in Fig.~\ref{fig:beamfractional}.  On propagation, the positions of vortices change significantly, and other vortex pairs appear, but the chain remains.  The infinite line of vortices is, of course, eventually lost in the low intensity regions of the beam tail, but we may say that the ``signature'' of Hilbert's Hotel still remains.  As already mentioned, this chain for a finite beam was already observed long ago \cite{jleymjp:njp:2004}, though not connected with the Hotel.

It is to be noted that the calculations presented here used the paraxial approximation inherent in the Fresnel diffraction formulas.  It is not clear at this point whether the existence of the infinite fractional vortex Hotel depends upon this approximation, and this will be investigated in future work. 

The results presented here suggest that the mathematics of infinite sets can manifest in surprising ways in optics.  They suggest that transfinite mathematics may be hidden in even more optical systems, particular those that have vortices present.

This research is supported by the U.S. Air Force Office of Scientific Research (USAFOSR) under Grant FA9550-13-1-0009.  The author would like to thank Professor M.V. Berry for insightful and helpful discussions.

\renewcommand{\baselinestretch}{1.0}
\tiny

\normalsize
\vspace{1.0cm}

\bibliographystyle{unsrt}
\bibliography{gburrefs}

\end{document}